\documentclass[review,authoryear,10pt]{elsarticle}
\usepackage [latin1]{inputenc}
\usepackage{amsmath}
\usepackage{graphicx}
\usepackage{caption}
\usepackage{setspace}
\usepackage{amssymb}
\usepackage{natbib}
\usepackage{multirow}
\usepackage{booktabs} 
\usepackage{epsfig}
\usepackage[english]{babel}
\usepackage{lscape}
\usepackage{bm} 

\usepackage{enumerate}
\usepackage[breaklinks=true]{hyperref} 
\usepackage{breakurl} 
\usepackage{balance}
\hypersetup{colorlinks=true,citecolor=blue, linkcolor=black,urlcolor=blue} 

\usepackage{color}

\usepackage[a4paper,top=2.5cm,bottom=2.5cm,left=2.5cm,right=2.5cm,bindingoffset=5mm]{geometry}

\begin{document}

\begin{frontmatter}



\title{Quantifying the distribution of editorial power and manuscript decision bias at the mega-journal PLOS ONE}
%
\author[1]{Alexander M. Petersen}
\ead{apetersen3@ucmerced.edu}
\address[1]{Ernest and Julio Gallo Management Program, School of Engineering, University of California, Merced, California 95343}

\begin{abstract}
We analyzed the longitudinal activity of nearly 7,000 editors at the mega-journal PLOS ONE over the 10-year period 2006-2015. 
Using the article-editor associations, we develop editor-specific measures of power, activity, article acceptance time, citation impact, and editorial renumeration (an analogue to self-citation). We observe remarkably high levels of power inequality among the PLOS ONE editors, with the top-10 editors responsible for 3,366 articles -- corresponding to 2.4\% of the 141,986 articles we analyzed; the Gini-index of this power distribution is 0.583, which is  comparable to some of the highest wealth-inequalities in the world.  Such high inequality levels suggest the presence of unintended incentives, which may reinforce unethical behavior in the form of decision-level biases at the editorial level. Due to the size and complexity associated with managing such a large mega-journal, our results indicate that  editors may become apathetic in judging the quality of articles and susceptible  to modes of power-driven misconduct.
We used the longitudinal dimension of editor activity to develop two panel regression models which test and verify the presence of editor-level bias. 
In both models we clustered the articles within each editor's profile and used editor fixed-effects to isolate the individual-level trends over time: in the first model we analyzed the citation impact of articles, and in the second model we modeled the decision time between an article being submitted and ultimately  accepted by the editor. 
We focused on  two variables that represent social factors that capture potential conflicts-of-interest: (i) we accounted for the social ties between editors and authors by developing a measure of repeat authorship among an editor's article set, and (ii) we accounted for the rate of citations directed towards the editor's own publications in the reference list of each article he/she oversaw. Our results indicate that these two factors play a significant role in the editorial decision process, pointing to the misuse of power. Moreover, these two effects appear to increase with editor age, which is consistent with behavioral studies concerning the evolution of  misbehavior and response to temptation in power-driven environments. And finally, we  analyze  ``editor renumeration'' -- the number of citations one might receive by adapting biases towards certain scientific peers as well as self-citations from scientific strangers.   By applying quantitative evaluation to the gatekeepers of scientific knowledge, we shed light on various issues crucial to  science policy, and in particular, the management of  large megajournals.
\end{abstract}


\end{frontmatter}


\section{Introduction} 
The emergence and rapid growth of megajournals  in the last decade \citep{Solomon_OpenAccess_2012,Binfield_megajournals_2013,Solomon_MegajournalSurvey_2014,Bjork_Megajournalgrowth_2015} represents a drastic industrial paradigm shift in the production of scientific knowledge. This paradigm shift   places pressure on several fundamental aspects of the  scientific endeavor. First, in addition to the  pay-to-publish model, the  personnel resources required to  referee the 50,000+ megajournal articles each year is quite substantial \citep{Binfield_megajournals_2013}. Second, the publication of these articles also stresses the individual cognitive capacity of scientists as well as the technological knowledge-storing capacity which is fundamental to the long-term need to be able to search, retrieve, and classify knowledge. For example,  over its first 6 years, PLOS ONE grew at an annual rate of 58\%, roughly 18 times larger than the net growth rate of scientific publication over the last half-century; in 2012,  the 23,468 articles published in PLoS ONE  represented approximately 0.1\% of science publications indexed by Thomson Reuter's Science Citation Index \citep{Pan_memory_2015}. And third,  megajournals have initiated a completely different model of managing the scientific publication process. In particular, PLOS ONE relies on thousands of acting scientists who comprise its editorial board, who simultaneously continue their role as research leaders. This dichotomy clearly establishes the conditions for conflicts-of-interest, whereby scientists must balance conflicting incentives arising from their distinct duties as authors and editors. 

As such, megajournals may be particularly susceptible to misconduct -- by authors and journal editors -- because oversight and careful monitoring of individual activities is truly challenging  in such a large and complex socio-economic system. This is, to some extent, supported by the lack of transparency in the review process, which has traditionally involved single or double blinds for the authors and reviewers, and also for the editors whose identity is oftentimes unknown even after the article is published.\footnote{However, unlike most journals, PLOS ONE provides the name of the editor overseeing each article, a crucial aspect which we  leverage in this study. } 
Particular to the review process of  science, which is ironically  obscured to protect the process itself, it is not difficult to imagine that misconduct may organically arise from the basic pursuit of internal (and external) power  \citep{Malhotra_Pursuit_2011} and the innate difficulty of avoiding temptation in decision-heavy endeavors \citep{Gino_Unable_2011}.  
As such, given the history of misconduct in science combined with  the relative ease in which information can be collected and made publicly available, it is important to develop transparent methods to quantitatively  monitor the activities of both authors and editors, so to ensure the integrity of the scientific process.

Here we provide an in-depth analysis of the  largest journal in the world, PLOS ONE, focusing on  editorial power and identifying the role of  social factors in their decision processes. To achieve this, we analyzed the entire article history over the 10-year period 2006-2015. By combining editor, author, article and citation level data for each publication, we constructed a large multi-variable longitudinal database centered around the 6,934 PLOS ONE editors. We use this data to  provide descriptive and panel regression analyses, which foster insight into a domain of science that has traditionally been undocumented since most journals do not make clear the editor-article association within the article. As such, to the best of our knowledge, this is the first study to develop editor-level measures for quantitatively evaluating an editor's history of editorial service. Moreover, these methods are the starting point for measuring individual and population-level shifts in editorial behavior over time.

The general hypothesis that we  test in this work is whether  bias exists in the editorial process. Identifying, with certainty, cases of scientific misconduct on the part of editors would require case-by-case investigation which is beyond the capacity of our data and methods. Thus, we refrain from orientating our study on particular editors, instead focusing on aggregate  patterns that nevertheless indicate that editorial behaviors shift over time as they gain power and, as a result, reach the limits of their temporal and cognitive capacity to pay close attention to each article they oversee.
This point is particularly important in the context of large megajournals, which are  typically electronic online-only and based on a  pay-to-publish model, meaning that the production process is primed for steady, and possibly overwhelming, growth.
Moreover, research indicates that when the acceptability of misconduct increases gradually,   that a ``slippery-slope effect'' \citep{Gino2009708} may emerge, and the spread of misconduct may be  inevitable even among individuals who initially had good intentions. 
It is likely that scientific actors are susceptible to such forces, because the information concerning the review process is tightly concealed. Ironically, instead of protecting the system, this lack of transparency may harbor the emergence of author and editor-level strategies for ``gaming the scientific system''. Given the increasing role of citation-based  evaluation -- of researchers and journals -- it  is entirely possible that misconduct may straddle both sides of the table.

In order to provide light on this  issue, we develop methods to quantitatively measure the author-editor relationship. In particular, we investigate two 
types of cooperative social ``back-scratching'': (i) editors making  biased decisions in the review process towards their close scientific peers, and (ii) authors enticing favorable decisions by providing renumeration in the form of citations directed to the editor's research. We leverage  the longitudinal aspects of editorial profiles to demonstrate for these two effects that, indeed, there are statistically significant trends in editor behavior at PLOS ONE that are consistent with power-driven bias and renumeration incentives.

\section{Literature review}\label{sec:liter_hyp}

This work contributes to several research streams focused on understanding the economics of science \citep{Stephan:2012}, the growth of  scientific  production \citep{Pan_memory_2015}, and  ethics issues arising from this growth paradigm \citep{petersen_quantitative_2014}. 
Our work also draws on the scientific community's efforts to  develop data-driven  models for the social processes underlying science \citep{ScharnhorstBoernerBesselaar2012,SciSciPolicyEditorial2016}.   
To this end, we employ a  data-driven approach that leverages the vast amounts of available publication metadata to expand our knowledge about science itself \citep{Evans_Metaknowledge_2011}. In  particular,  our effort to capture the obscure yet fundamental interactions between individuals -- here the editor-author relation -- which builds on concepts and methods from social science is only  possible due to recent simultaneous advances in computing, online data availability   \citep{Lazer2009}.

As science continues to expand, it is important to frame introspective 'science of science' analysis around important questions of  science policy \citep{fealing_science_2011,Stephan:2012,SciSciPolicyEditorial2016} -- e.g. how to increase the efficiency of scientific discovery \citep{Rzhetsky24112015} and  improve the evaluation and sustainability of scientific careers in an increasingly metrics-oriented system \citep{TheMetricTide}.  Indeed, scientists should be developing ideas to improve the publication system to the same degree that they are  rethinking the scientific  funding system \citep{Bollen_novelfudningsystem_2016}. If not, it is possible that the system may become susceptible growth of inequality \citep{ScienceInequality,CumAdvEPJDS}, corruption, and unintended consequences arising from lack of oversight, which are  common flaws of  complex human-oriented systems. In addition to these prominent flaws, the scientific process is also characterized by   subtle innate biases -- such as the editorial and peer review bias in the reporting of positive scientific results, as apposed to  null results   \citep{Kravitz_Editorial_2010} 

This study also provides a timely quantitative insight of the largest among the rapidly growing ecosystem of megajournals, which feature journals that are publisher, society, and research-area specific, such as  Heliyon (Elsevier), Springer Plus (Springer), Scientific Reports (Nature), Royal Society Open Sciences, IEEE Access, PeerJ, SAGE Open \citep{Solomon_OpenAccess_2012,Binfield_megajournals_2013,Solomon_MegajournalSurvey_2014,Bjork_Megajournalgrowth_2015}.
And finally, our analysis also contributes to recent research on manuscript decision timescales  \citep{Powell_Wating_2011,Sugimoto_AcceptanceRates_2013} by identifying social factors that can explain the wide range in acceptance times.


\section{Data and Methodology}\label{sec:data_method}

\subsection{PLOS ONE article data}\label{subsec:data}
We gathered the citation information for all PLOS ONE publications from the 
\href{http://thomsonreuters.com/thomson-reuters-web-of-science/}{Thomson Reuters Web of Science (TRWOS)} Core Collection. 
From this data we obtained a master list of the unique digital object identifier, $DOI_{A}$, as well as the  number of authors, $k_{A}$, a list of their surnames and first-middle name initials, and the number of citations, $c_{A}$, at the time of the data download (census)  date on December 3, 2016. 

We then used each $DOI_{A}$ to access the corresponding online XML version of each article at \href{http://journals.plos.org/plosone}{PLOS ONE} by visiting the  unique web address given by each   ``http://journals.plos.org/plosone/article?id='' + ``$DOI_{A}$'' string combination. 

\subsection{Article and editor measures}
In our study, the principal unit of analysis is a PLOS ONE editor, which we denote by the index  $E$. Thus, for each $E$ we collected the corresponding group of $N_{E}$ articles over which he/she has served as editor. This Editor-Article association is publicly available in both the  published electronic article as well as on the article webpage, appearing in the article abstract and author information byline.  To maintain context among the variables we define in our analysis, quantities that are mostly article-specific are denoted by the index $A$, those that are mostly editor specific are denoted by the index $E$, and quantities that are properties of both are indexed as $x_{A,E}$.

 Embedded in the XML file for each article are various  editor, coauthor, and article metadata which we extracted from the webpage of each $A$ and then aggregated  for each $E$. All together, the entire database for the 10-year period 2006-2015 is comprised of  141,986 articles and 6,934 editors. In both of our panel regression models, we refine this dataset to the  3749 editors with $N_{E}\geq10$ articles to reduce small sample noise at the editor level, resulting in 128,734 articles. From these articles and their editors we define the following quantities:
\begin{enumerate} 
\item The net editorial activity, $N_{E}$, is the number of articles overseen by editor $E$ over the total  editor service  period,   $L_{E}$, which is the   number of days between an editor's first and  last article --  through the end of 2015. 
\item The article acceptance time, $\Delta_{A}$, is the number of days between the submission and acceptance of article $A$. Note that this duration does not include the time interval between acceptance and publication, as factors external to the editorial process could affect this process, its timeline, and thus its ultimate duration.
\item The editorial turnover time, $d_{E}=L_{E}/N_{E}$, is the mean number of days between two articles overseen by editor $E$  published in PLOS ONE. 
\item The editorial mean acceptance time, $\Delta_{E}$, is the mean $\Delta_{A}$, which is calculated within his/her article group. Likewise, we measured the variability in  $\Delta_{A}$  using  the coefficient of variation,  $cov_{E} = \sigma_{E}[\Delta_{A}]/\Delta_{E}$,  calculated within each editor's article subset (where $\sigma[...]$ denotes the standard deviation); see Fig. \ref{figure:Figure-S2}(A).
\item The ``citation renumeration'' $C_{A}$ is the total number of references  that cite the editor's research  among the articles he/she edited. This number is calculated by going through the reference list of each article, and identifying publications that include the  editor's last name and first-name initial among the authors. Likewise, the editor citation rate, $f_{A}$, is the fraction of the total references on a given article that cite his/her work. 
\item The editor's PLOS ONE service age, $\tau \equiv \tau_{A,E}$, is the time difference between the  acceptance date of the first accepted article of editor $E$ and the acceptance date of the acceptance date of article $A$, measured in years. 
\end{enumerate}
Figure \ref{figure:EditorDataSummary} shows the probability distribution for several of these important quantities, with article-level statistics shown in blue, and editorial-level statistics shown in red throughout the remainder of the analysis.

\subsection{Article subject area classification}\label{subsec:method}
It is well known that citation rates are affected by discipline-dependent factors. Indeed PLOS ONE is comprised of articles from a range of disciplines, and  is classified by TRWOS as a ``Multidisciplinary'' journal. Thus, we were careful not to blindly pool the citation impact measures from all articles together. Instead, we methodically separated the articles into subsets, so that the relative citation difference between two articles less biased by (sub)disciplinary factors such as research community size and innovativeness, but rather,  is an estimate of  differences in research quality and scientific impact. 

We grouped the  articles into 6 refined core subject areas (SA) based on the internal PLOS ONE classification \href{http://journals.plos.org/plosone/s/help-using-this-site#loc-subject-areas}{subject area classification system} derived from a controlled thesaurus of more than 10,000 keywords. 
To be specific, we started with the keywords appearing on the webpage  of  article $A$. Nearly all articles have 8  keywords per article, with only a handful of articles containing less than 8. 
PLOS ONE also has an article-classification scheme which is used to group articles for comparing article visibility. The 2-level classifications are accessible in the ``page-views'' applet on each article's ``Metrics'' page, for example the article with DOI 10.1371/journal.pone.0000112 is classified primarily as ``Biology and life sciences'', with 3 subclassifications (Evolutionary Biology, Genetics, and Population Biology).   In all, the 10 top-level classifications are:
(i) Biology and life sciences,
(ii) Medicine and health sciences,
(iii) Physical sciences,
(iv) People and places,
(v) Social sciences,
(vi) Engineering and technology,
(vii) Computer and information sciences,
(viii) Ecology and environmental sciences,
(ix) Earth sciences,
(x) Science policy.
We collected these top-level classifications from each article's ``Metrics'' page, and are ordered here according to their frequency among PLOS ONE articles.

 We then used the statistics of the bipartite association between keywords and top-level classifications on each article to establish a vector of weights corresponding to the 10 top-level PLOS ONE classifications in such a way that we could more precisely identify an article with a single classification.
More specifically, for each article, we identified the principal SA as the one for which the individual keywords contributed the most weight. Take again the article  DOI:10.1371/journal.pone.0000112 with the 8 keywords ``Chromosome 4'', ``Genetic loci'', ``Centromeres'', ``X chromosomes'', ``Population genetics'', ``Chromosomes'', ``Sex chromosomes'', ``Alleles''. As one might expect, these article keywords give the largest weight to  SAs (i)``Biology and life sciences'' and (ii)``Medicine and health sciences''. 

Applying this method to all articles, we found that the most common first and second ranked classification across all PLOS ONE articles are indeed (i) and (ii).  Figure \ref{figure:Figure-S1}(A) shows the SA count histogram for all PLOS ONE articles, with 123,750 (87.1\% of all articles) having ``Biology and life sciences'' as the principal classification, and none having ``Science policy'' as the principal classification. Contrariwise, only 17 articles had ``Earth sciences'' as the principal classification. To account for the fact that the majority of the keywords in the PLOS ONE thesaurus are related to (i)  and (ii), leading to the disparity in the principal classification, we created an exception rule in order to better account for the second-ranked classification. First, if the principal classification was (i), then we instead used the second-ranked classification as the principle classification. This rule helped to classify more publications for SA (iii)-(ix), as demonstrated by the second count histogram shown in Fig. \ref{figure:Figure-S1}(B). As one final step to condense the SA classifications, we joined the groups (iv) and (v), (vi) and (vii), and (viii) and (ix), since there is considerable  overlap between these classification groups. Thus, Fig. \ref{figure:Figure-S1}(C) shows the final refined distribution of articles across the 6 refined SA used in our analysis: the smallest refined SA is 6/7 with 533 articles and the second-smallest is 4/5 with 1839 articles over 2006-2015; the remaining refined SA are comprised of $8000$ or more articles over the 10-year period. In what follows, we define the variable $s=1...6$ as an index for these refined SA; we use $s$ as a subset index in defining normalized citations and dummy variables in our regression analysis.\\

\subsection{Normalization of citation counts to account for citation inflation and SA-level factors}
Comparing the raw  citation counts of articles from different years $t$  and disciplines $s$  is a common challenge in science of science research. This difficulty arises due to three principal statistical biases: variation in publication rates across discipline, censoring bias and citation inflation. The first refers to the fact that larger disciplines, e.g. Biology and life sciences, produce more publications, and hence, more citations than smaller disciplines such as Earth sciences. The second  bias refers to the fact that older publications have had more time to accrue citations than newer ones.  And the third bias refers to the fact that more citations are produced over time as a product of increasing publication rates and reference list lengths, leading to a significant  inflation in the relative value of citations. By way of example,  a recent study demonstrated that the total number of references produced by all scientific articles is growing by 5.6\% annually, and hence doubling every 12.4 years \citep{Pan_memory_2015}.   

To address these three measurement problems, we map the raw citation count  ${c}^{s}_{A,t}$ of a given article -- measured at the TRWOS census date $Y=12/03/2016$ --    to a normalized value 
\begin{equation}
z^{s}_{A} \equiv  \frac{\ln (1+{c}^{s}_{A,t}) -   \langle \ln (1+{c}^{s}_{t}) \rangle}{ \sigma[ \ln (1+{c}^{s}_{t})]} \ ,
\label{zn}
 \end{equation} 
where  the mean, $\langle \ln (1+{c}^{s}_{t}) \rangle$, and   the standard deviation, $\sigma[ \ln (1+{c}^{s}_{t})]$, are calculated only over  publications from  the same year $t$ and  refined subject area $s$. The constant  1 is added to each citation count in order to avoid the problem of uncited articles and does not affect the results. 

By analyzing the logarithm of the citation count, this normalization leverages the universal log-normal statistics of citation distributions \citep{UnivCite}. Moreover, by rescaling the logarithm by the standard deviation, the underlying inflationary bias has been removed from $z$, i.e. detrended so to permit cross-era comparison. As such, $z$ is particularly well-suited for regression analysis, as recently demonstrated in longitudinal analyses of cumulative advantage \citep{CumAdvEPJDS} and collaboration \citep{petersen_quantifying_2015} within researcher careers. Figure  \ref{figure:Figure-S1}(B) confirms that the probability distributions $P(z|s,t)$  are all approximately normally distributed, and thus  sufficiently  time invariant for the purposes of our analysis, for each subject area and year -- with the exception of 2015 publications, which we omit from our first regression analysis where $z$ is the dependent variable. 

\subsection{Repeat authors}
In order to investigate the impact of social ties on editorial decision processes, we analyzed the set of $N_{k}$ authors appearing within the article set of a given editor. That is, for each article we recorded the last name and first initial of each of the $k_{A}$ coauthors and tallied the number of articles $A_{E,k}$ for a given surname + first-initial combination -- for each  editor. Because of the name disambiguation problem, it is difficult to distinguish authors with the same name, especially for authors with extremely common surnames. Thus, we removed from our analysis those authors with common surnames (e.g. Xie, Yang, Adams, Johnson), using the Editor name list to determine which surnames appear with significant frequency that might significantly contribute to false-positive union of coauthor counts. We describe this procedure in the Supplementary Information Section \ref{NameDisamb}, where we also provide the full list of surnames which we ignored in our  analysis.

After tallying the author names for each $E$ we obtained a ranked list of those coauthors. We investigated the distribution of the rank-coauthor profile within an editor's article set, and found that the distribution $P(A_{E,k})$ decays quasi-binomially, but with  deviations in the tail. Also, as expected, the  maximum value Max$[A_{E,k}]$  depends to a large degree on $N_{E}$. Thus, unlike the rank-coauthor distribution within a given researcher's publication profile, which is well-fit by a discrete exponential distribution and characterized by a subset of ``super-ties'' representing extremely strong collaboration partners \citep{petersen_quantifying_2015}, the editor-author distribution is not characterized by such obviously strong social ties. 

Nevertheless, since the purpose of this method is to identify a set of articles which may have been influenced by  external social ties with the editor,  the ``repeat coauthor'' criteria we converged upon leverages the fact that many top authors within an editor set tend to publish together. As such, we found that the best method for identifying a reasonably-sized subset of articles with repeat ties was to merely tag the authors with 2 or more articles within a given editor's article set -- i.e. repeat authors. 

 Applying this method, we counted the number of repeat authors per editor, $K2_{E}$, and we then tagged each of the articles  including those authors using the indicator variable $R_{A,E}=1$. As such, the articles with $R_{A,E}=1$ represent 13.9\% of all articles. Figure  \ref{figure:Figure-S2}(B) shows that, within editor profiles,   each editor has  on average 5.2 repeat authors. Likewise, on a per-article basis, Fig. \ref{figure:Figure-S2}(C) shows that   on average 11\% of an editor's articles have $R_{A,E}=1$ (the median value per editor is  0.1), however the distribution of this fraction $\rho_{E}$ is  skewed with 10\% of editors having 26\% or more of their articles with $R_{A,E}=1$. In all, this ``repeat author'' method  identifies a sufficiently large number of articles with $R_{A,E}=1$  (the rest having indicator value $R_{A,E}=0$), such that in the following sections we can use this binary classification to identify and quantify possible obscure social factors that may reasonably affect editorial decision processes.

\section{Results}\label{sec:results}

\subsection{Editorial power distribution}
Our analysis reveals an extremely  wide range of editorial power that exists within a single journal. 
For example, ranking editors by  $N_{E}$, Fig. \ref{figure:EditorDataSummary}(A)  shows  the  top-100 editors, who  collectively oversaw 12.2\% of the  total 141,986 articles over the 10-year period since the inception of the journal. Moreover, the distribution $P_{N_{A}}$  calculated across all editors is extremely right-skewed, showing that most editors have served on just a few articles, 50\% have served on 11 or less articles,  while the top editor Vladimir Uversky has served on roughly 27 times ($557/20.5$) as many articles as the average editor.

To further demonstrate this power inequality, the  Lorenz curve in Fig. \ref{figure:EditorDataSummary}(C) shows the cumulative fraction of all articles edited by a given percentile: the bottom 25\% of editors oversaw just 3\% of the total 141,986 articles; the middle 65\% of editors oversaw 55\%; the top 10\% of editors  (693 editors) oversaw 42\%; the top-10 editors oversaw 2.4\% of the total articles. The Gini-index, representing the area between the Lorenz curve and the diagonal line, is relatively high, comparable to some of the highest wealth inequalities in the world, e.g. in Honduras. 

\subsection{Editorial activity distribution}
\begin{figure*}
\centering{\includegraphics[width=.99\textwidth]{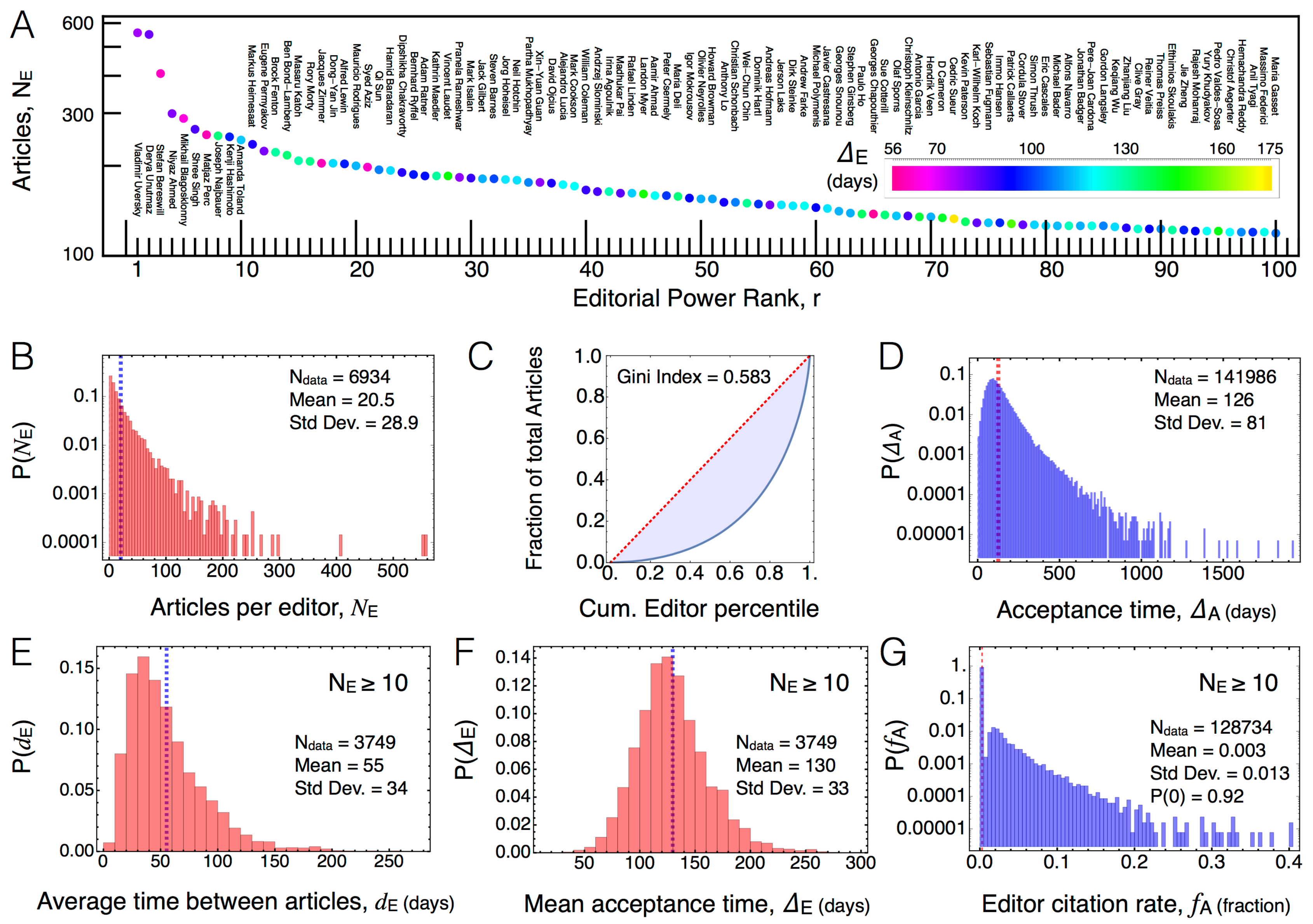}}
\caption{\label{figure:EditorDataSummary}  {\bf The distribution of editorial power, activity, and citation renumearation at PLOS ONE}.   (A) The top-100 most active editors ranked according to $N_{E}$. Circle color indicates the editor's mean time to acceptance, $\Delta_{E}$ (days); green-yellow values are above the population mean of 130 days (see panel F); blue-magenta values are significantly below the population mean, and are typical of the top-10 editors. (B) The  right-skewed distribution of $N_{E}$ across nearly 7 thousand editors.  (C) The Lorenz curve quantifies the cumulative fraction of all articles edited by a given percentile: the bottom 25\% of editors oversaw just 3\%  whereas the  the top 10\% of editors  (693 editors) oversaw 42\% of the total 141,986 articles. The Gini-index calculated from the  distribution of $N_{E}$ is 0.583. (D) The   distribution of the number of days between an article was received and accepted for publication (i.e. not including the time between acceptance and publication). (E) The distribution of the turnover time (or inverse activity) defined as the average number of days between articles accepted by the same editor. (F) The distribution of the mean number of days  to accept an article calculated for each editor; comparable with panel D. (G) The distribution of $f_{A}$, the fraction of the references in a given article that cite other papers that include the editor as a coauthor: 92\% of papers have $f_{A}$=0, but there is an extremely long tail. In panels (E-G) we only included data for the 3,749 editors with $N_{E}\geq 10$ articles in order to reduce the fluctuations due to small sample size; vertical dashed lines indicate distribution mean.}
\end{figure*}

The activity of editors is  right-skewed, which is partially attributable to the  skew in the distribution of acceptance times $P(\Delta_{A})$. Figure \ref{figure:EditorDataSummary}(E) shows that the mean time between articles accepted is on average roughly 55 days, about half as long as the mean $\Delta_{A}$. This would suggest that most editors are handling roughly two articles at a time, which is not unreasonable. Moreover, comparing  panels \ref{figure:EditorDataSummary}(D,F), the distribution $P(\Delta_{E})$ of   the mean number of days  to accept an article after it was received within an editor's article subset is comparable in mean -- but not standard deviation -- to the article-level distribution. One might assume that this variation is purely related to $N_{E}$, however the color coding of editors in Figure \ref{figure:EditorDataSummary}(A) by their individual $\Delta_{E}$ values indicates that among the top-100 editors there is an extremely wide range of accepted article turnover  levels. It appears as though the extremely high volume of articles overseen by the top-10 editors is partially due to their  fastest acceptance times, with several averaging just around 2 months per article. 

\subsection{Panel data regressions with observations clustered on Editor profiles}
In what follows, we aim to explain the variation in two types of outcomes -- the article's scientific impact and its speed through the review process --  using editor and article-level control variables. Of principal interest among the covariates are those representing social aspects of the editorial review process. Namely, we shall focus on the variation due to repeat authors, as indicated by the binary variable  $R_{A,E}$. Moreover, we shall also focus on potential evidence of editor renumeration using the rate of references directed at the editor's publications, $f_{A}$.  For example, Fig. \ref{figure:EditorDataSummary}(G) shows the distribution $P(f_{A})$, which indicates that 92\% of articles do not have any references that cite the editor's work. Nevertheless, among the remaining 8\% of articles with $f_{A}>0$, there is a wide range, with the average value $\langle f_{A} | f>0 \rangle = 0.036$ corresponding to roughly one in every 30 references citing the editor's work. And finally, the longitudinal component is rather important, as variation across each editor's service, captured by $\tau_{E}$, may indicate shifts in behavior reflecting increased workload, apathy, and possibly even new variations of scientific misconduct. 

\subsubsection{Model I: Article citation impact, $z_{A}$}
\label{subectionmodel1}
In this first model, we ask the question: does the scientific quality of the article's overseen by a given editor change over time? If so, what are the possible explanations? Indeed, similar approaches have been used to demonstrate that within researcher profiles, indeed there is a negative trend in the scientific impact of a researcher's publications the further on in his/her career \citep{CumAdvEPJDS,petersen_quantifying_2015} with various potential  individual and social mechanisms that may be responsible for this observed trend. 

 Thus, the dependent variable is the normalized citation impact of an article $z^{s}_{A}$  in subject area $s$. By matching each article to its editor $E$, we capture the longitudinal dimension quantified by  $\tau$, the number of years into  his/her editorship at PLOS ONE. 
 Thus, by sequencing the $A$ by $\tau$, we are able to use a panel regression framework including editor  fixed-effects ($\beta_{i,0}$)  to control for time-invariant individual-level characteristics.  Thus, this model appropriately captures within-career trends. 
 For this panel analysis we exclude articles from 2015, since our analysis of $z^{s}_{A}$ in the bottom row of Fig. \ref{figure:Figure-S1}(B) indicates that these articles have not had enough time to sufficiently  converged to the baseline  Normal $N(0,1)$ distribution.
 Thus, by considering only those editors with $N_{A}\geq 10$, this additional threshold reduces the dataset from 128,734  to 102,741 articles (observations).
 
 The specification of our linear fixed-effects model is given by
\begin{equation}
z^{s}_{A}=  \beta_{E,0} + \beta_{k} \ln k_{A}  + \beta_{\Delta} \ln \Delta_{A}  + \beta_{\tau} \ln \tau_{A,E}  +  \beta_{R}  R_{A,E} + D_{s} + D_{t}    + \epsilon_{A,E} \ .
\label{Eqn1}
\end{equation}
The results of this basic  model estimates are shown in the first two columns of  Table \ref{table:reg1}, where the second column corresponds to the standardized (beta coefficient) coefficients

The article-level variable $k_{A}$ controls for team-size effects, and is incorporated in logarithm since the distribution of authors per publication is right-skewed and approximately log-normal in various team-oriented disciplines \citep{petersen_quantitative_2014}. Along these lines, we also include subject area as well as publication year dummies variables to further control for cross-disciplinary and temporal variation in the explanatory variables. 

The first covariate of interest is $\Delta_{A}$, the amount of time it took for the article to be accepted. 
The model indicates that publications that are under review for a longer time tend to have lower citation impact ($\beta_{\Delta}<0 ; p<0.000$). This is consistent with the assumption that articles that fail to signal their novelty and/or scientific contribution may require more deliberation time between the authors, the reviewers, and the editor. 

The principal explanatory variable of interest, $\tau_{E}$, the duration of editorial service at PLOS ONE at the time of acceptance of the article, was the strongest variable in explaining $z_{E}$, with standardized coefficient $\hat{\beta_{\tau}} = -0.143$ $(p<0.000$).
This result is consistent with two other studies of longitudinal citation patterns within careers \citep{CumAdvEPJDS,petersen_quantifying_2015}, and suggests  that editorial behavior may shift across his/her career.  

We also incorporated the repeat author indicator $R_{A,E}$, which has a significant positive coefficient $\beta_{R}>0$, possibly due to the fact that repeat coauthors are likely to be more experienced and more prominent within their scientific community. In order to identify additional  signatures of editorial bias, we estimated two additional models, the first  including an additional interaction term $R_{A,E}\times \ln \tau_{E}$ and the second including an additional triple-interaction term $T_{10,E}\times R_{A,E}\times \ln \tau_{E}$, where $T_{10,E}=1$ if the editor is ranked in the top-10  according to $N_{E}$ and 0 otherwise. 
The former double-interaction term captures the potential combined effect of editor age and social ties while the latter triple-interaction term captures the additional effect of being extremely prominent editor at PLOS ONE. 
The model estimates for the  triple-interaction are shown in the fourth column of Table \ref{table:reg1}, and indicate that articles with $R_{A,E}=1$ have decreasing impact for larger $\tau$ ($\beta_{T\times \ln \tau}= -0.025$; p=0.028). Moreover, we observe an additional negative effect related to the combination of top-10 editors who accept articles with $R_{A,E}=1$ for larger $\tau$ ($\beta_{T\times R \times \ln \tau} = -0.103$; p=0.017). 

\begin{figure*}
\centering{\includegraphics[width=.99\textwidth]{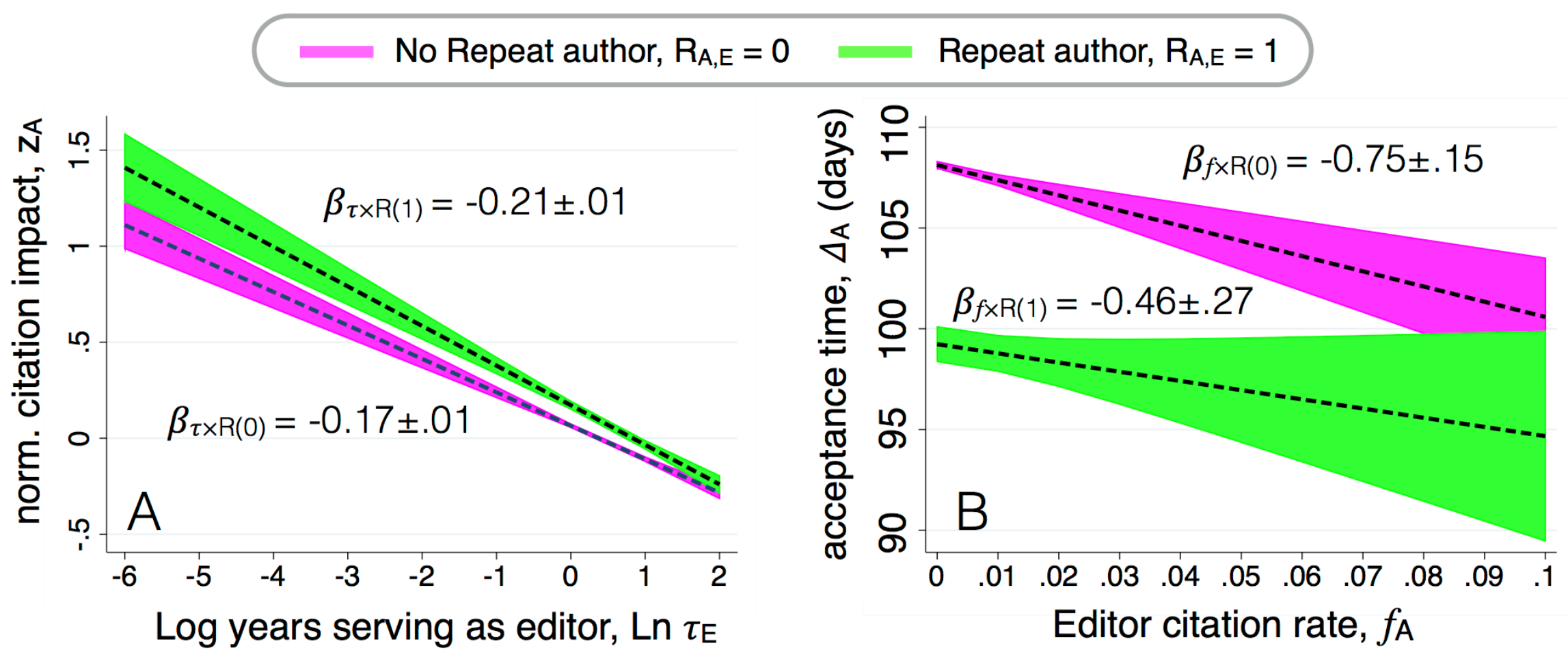}}
\caption{\label{figure:ModelMarginalEffects}  {\bf Quantifying editors' diminishing quality judgment and susceptibility to self-citation.} Shown are linear predictions represented as point estimates  with 95\% confidence intervals  using two fixed-effects models: (A) the normalized citation impact model specified in Eq.~\ref{Eqn1}  and (B) the acceptance-time model specified in  Eq.~\ref{Eqn2}.   (A) The marginal effect of editorial longevity ($\ln \tau_{E}$) on the normalized citation impact ($z_{A}$) of the articles he/she accepts,  including an interaction term $R \times \ln \tau$ to distinguish between those articles with $R=1$ (with repeat authors) and $R=0$ (no repeat authors). Both interaction coefficients are negative and significant at the $p<0.000$ level; the difference in the coefficients is significant at the $p\leq 0.005$ level.
 (B) The marginal effect of citing the editor's publications ($f_{A}$) on the time the editor takes to accept an article $\Delta_{A}$, including an interaction term $R\times f$ to distinguish between the articles with $R=1$ and $R=0$. 
Both  interaction coefficients are negative, however  due to small sample size for the $R=1$ cases, only $\beta_{f\times R(0)}$ ($p<0.000$)  is significant, with $\beta_{f\times R(1)}$ only significant at the $p=0.096$ level. 
Shaded interval indicates the  95\% confidence interval calculated using the delta method with all covariates evaluated at their mean values. }
\end{figure*}

Figure \ref{figure:ModelMarginalEffects}(A) captures the marginal effect of editor service age $\tau$ on $z_{A}$ for the two scenarios for $R_{A,E}=0,1$. Indeed, both trends are negative, however the marginal effect for $R_{A,E}=1$ is even more negative such that by $\ln \tau=1$, corresponding to roughly 3 years of service, the  characteristic article citation impact is below average with no significant difference from the articles with $R_{A,E}=0$.

\subsubsection{Model II: Article acceptance time, $\Delta_{E}$}
\label{subectionmodel2}

The average PLOS One article takes 126 days from being officially received and processed by the editor, reviewed (possibly over several rounds), and finally accepted. This characteristic timescale is higher than the global average which was recently estimated to be roughly 100 days, with only slight variations across journals when disaggregated by impact factors  \citep{Powell_Wating_2011}.
However, there is great variation in the acceptance time $\Delta_{A}$, demonstrated by the distribution $P(\Delta_{A})$ in Fig. \ref{figure:EditorDataSummary}(D). This variation is highlighted by two remarkable extremes -- we  observed one publication with $\Delta_{A}=0$ (DOI:10.1371/journal.pone.0031292) and one publication  (DOI:10.1371/journal.pone.0028904) with $\Delta_{A}=1927$ days -- taking more than 5 years to finally be accepted! Moreover, we find that 0.43\% of  articles are received and accepted within 7 days.

Thus, in this second model, we ask the question: what factors may explain the wide range of acceptance times observed across all articles and even within the profiles of individual editors? Can we interpret the results in terms of the renumeration expectations for editorial service? The short preemptive answer is 'Yes' -- in the next section we shall further illustrate  the significant magnitude of renumeration an editor might be able to achieve according to unintended incentives in science aligned with the pursuit of  individual citation prestige.

For this panel analysis we use data for editors with $N_{A}\geq 10$, corresponding to 128,734  articles (observations).
 The specification of our linear fixed-effects model is given by
\begin{equation}
 \ln \Delta_{A} =  \beta_{E,0} + \beta_{z} \ln z_{A} + \beta_{k} \ln k_{A}  + \beta_{f}  f_{A}  + \beta_{\tau} \ln \tau_{A,E} +  \beta_{R}  R_{A,E} + D_{s} + D_{t}    + \epsilon_{A,E} \ .
\label{Eqn2}
\end{equation}
The results from our model parameter estimates are shown in Table \ref{table:reg2} along with their standardized (beta coefficient) counterparts.

In addition to the covariates used in the citation impact model, we also included the citation rate to the editor's articles ($f_{A}$, which did not show a significant effect in the citation model, nor would it have a logical mechanism,  and hence we  excluded it from that model). Consistent with the first model, higher impact articles tend to get accepted more quickly ($\hat{\beta}_{z}=-0.0343$; $p<0.000$), as their relative quality may be more easy to identify and accept faster. Also, the more coauthors on the article, the longer the articles tended to take in order to be accepted, in line with expected increasing coordination costs in assembling and submitting referee revisions  in large team endeavors ($\hat{\beta}_{k}=0.0297$; $p<0.000$). There was also a significant positive relation between editor service age and the acceptance time, in line with the increasing time demands as an editor becomes busier within the journal compounded by additional external activities ($\hat{\beta}_{\tau}=0.137$; $p<0.000$).

This  model also indicates a negative relation between editor citations and acceptance time ($\beta_{f}<0$), albeit the weaker in relative magnitude  as indicated by its standardized coefficient ($\hat{\beta}_{f}= -0.009$; $p<0.000$). Nevertheless, this significant negative relation again points to an undocumented type of editor bias in the handling of scientific manuscripts, pointing to the renumeration value of a self-citations. 

Again, of principal interest is the relation with repeat authors as an indicator of social factors. Indeed, the model indicates a significant reduction in  $\Delta_{A}$ for repeat authors ($\beta_{R}=-0.0878$; p< 0.000). In real terms, for the average article, this effect corresponds to roughly a  $\langle \Delta_{A} \rangle (1 - \exp[-0.0878]) =$ 10-day decrease in acceptance time related to $R_{A,E}=1$ alone. In order to further identify differences in $\Delta_{A}$ arising from vanity incentives, we added an interaction term $f_{A}\times R_{A,E}$ to the model specified in Eq. \ref{Eqn2}, in order to distinguish the relation of $f$ between  articles with $R=1$ and $R=0$. Figure \ref{figure:ModelMarginalEffects}(B) captures the marginal effect of $f$ on $\Delta_{A}$, with articles with $R_{A,E}=0$ showing a statistically significant negative relation $(\beta_{f\times R=0}=-0.75$; $p<0.000$, which, if leaning towards a more pessimistic interpretation, would indicate that ``{\it you must pay more the first time}.''

\subsection{Unintended incentives associated with Editorial board service}
Above all, it is impossible to use the data at hand to know the exact context of each of the references citing PLOS ONE editors' work. Thus, it is best to assume that the majority of these editor citations follow the same intent purposes of any other citation reference. 
Nevertheless, motivated by the results of the latter  regression model, it is clear that citing the editor -- loosely interpretable as an  unsolicited  renumeration for editorial services -- may actually work towards enticing a faster (positive) decision. But to what extent would an editor gain from quietly ``playing this market''? 

Indeed, in the previous sections, we found evidence that editors may treat repeat authors slightly differently, by lowering their standards of quality judgement and providing faster decision times to the (groups of) authors who an editor repeatedly serves, and possibly has connections to, within the larger academic community. Thus, in this section we seek to estimate the potential net ``renumeration'' that these repeat authors could effect by citing an editor's work. Of course, knowledge of the editor's identity is single-blinded at the point of submission, however  it is not impossible that an editor might communicate externally to the authors that he/she is overlooking their manuscript, or that the author's might informally contact a suspected editor. 

To this end,  by leveraging the size of the PLOS ONE dataset, we looked for small  but   measurable differences in the citation rate to editors conditional on the article including or not including repeat authors ($R_{A,E}=$ 1, 0, respectively). 
To be specific, for each editor we collected the set of $N_{E,R=1}$ articles with $R_{A,E}=1$ and counted the total number of references $C_{R=1}$ and also the number of those references citing the editor's work, $C_{E,R=1}$.  Similarly, for the set of $N_{E,R=0}=N_{E}-N_{E,R=1}$ articles with $R_{A,E}=0$, we also calculated $C_{R=0}$ and $C_{E,R=0}$. Thus, the total number of references from all articles overseen by an editor is simply $T_{E}=C_{R=0}+C_{R=1}$,  and the total number of citation received by the editor, independent of $R$, is $C_{E}= C_{E,R=1}+C_{E,R=0}= f_{E} T_{E}$.

\begin{figure*}
\centering{\includegraphics[width=.5\textwidth]{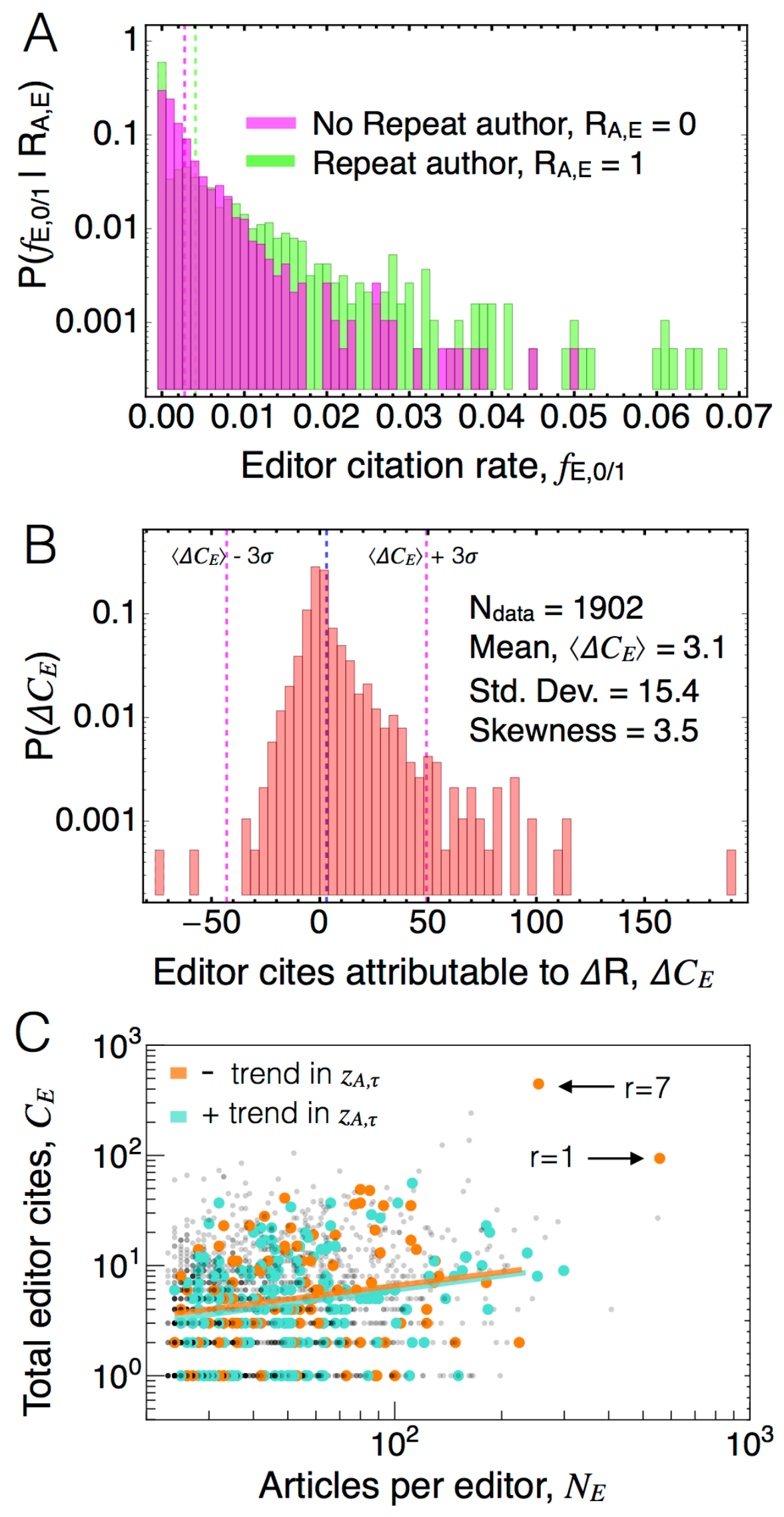}}
\caption{\label{figure:Editorcites}  {\bf Estimating the citation renumeration for editorial service}.   
(A) Conditional citation rate distributions for the articles without any repeat authors $P(f_{E} \vert R=0)$ (magenta), and for the articles with repeat authors  $P(f_{E} \vert R=1)$ (green). The mean values (indicated by the vertical dashed lines) are $\langle f_{E,0}\rangle = 0.0028$ and$\langle f_{E,1}\rangle = 0.0041$. Statistical tests for the difference in means (T-test), difference in median (Mann-Whitney test), and difference in distribution (Kolmogorov-Smirnov test) all reject the null hypothesis that the mean, median, and distributions are equal at the $p<0.000$ level.  
(B) Distribution of the excess editor cites due repeat author bias, estimated using the empirical editor-specific  difference   $\Delta C_{E} \propto (f_{E,1}-f_{E,0})$ for each author (see Eq. \ref{DCE}). 
The $\langle \Delta C_{E} \rangle \pm 3 \sigma_{\Delta C}$ confidence intervals are indicated  by vertical dashed magenta lines. The asymmetry in the tails of the distribution are evident when considering the outliers, with 39 observations in the right tail and only 2 in the left tail.
(C) Scatter plot of editor power ($N_{E}$), citation renumeration ($C_{E}$) and color indicating the trend in $z_{A}$ among the articles he/she accepted over time: editors with  significant positive (negative) trend in $z$ are colored cyan (orange), and  editors with no significant trend (p-value of the regression $p<0.1$) are colored grey. The two arrows indicate two  outliers characterized by high $N_{E}$, $C_{E}$, and negative trend in $z_{A}$, showing their  rank identity  in Fig \ref{figure:EditorDataSummary}(A). Solid lines show the regression best-fit between $\log_{10} N_{E}$ and $\log_{10}(C_{E})$, ignoring the two data points with the largest $N_{E}$ values within each subset; the fits are nearly identical. Each panel is calculated for editors with $N_{E}\geq 20$. }
\end{figure*}

We then define the conditional editor citation rates $f_{E,1}=C_{E,R=1}/C_{R=1}$  and $f_{E,0}=C_{E,R=0}/C_{R=0}$ and plot their  distributions $P(f_{E} \vert R_{A,E}=0,1)$ in Fig. \ref{figure:Editorcites}(A). The distribution calculated for publications featuring repeat authors ($R=1$) shows an excess in the extreme right tail, suggesting that these relatively high levels of editor citation rates may correspond to ``enticing  backscratching''.

In order to further examine signatures of renumeration, we calculated the expected number of citations that an editor might gain due to the differences in citing behavior of repeat versus non-repeat authors.  
We measure this difference as 
\begin{equation}
\Delta C_{E} = (f_{E,1}-f_{E,0}) T_{E} ,
\label{DCE}
\end{equation}
which should be equal to 0 for those editors who are completely unbiased with respect to $R$. Of course, there are fluctuations due to the finite sample size $N_{E}$. Figure \ref{figure:Editorcites}(B) shows the probability distribution $P(\Delta C_{E})$ is approximately normally distributed,  centered around the mean value $\langle \Delta C_{E} \rangle =3.1$ and standard deviation $\sigma_{\Delta C}=15.4$. However, the distribution deviates from the symmetry of the normal distribution, showing significant right skew (skewness = 3.5) representing the excess number of editors with relatively large and positive  number of citations attributable to differences in the citation rates for $R=1$ versus $R=0$. 
Remarkably, we count 39 editors with $\Delta C_{E} > (\langle \Delta C_{E} \rangle + 3 \sigma_{\Delta C})$, representing 2\% of the 1902 editors we analyzed with $N_{E}\geq 20$; we counted only 2 editors with $\Delta C_{E} < (\langle \Delta C_{E} \rangle - 3 \sigma_{\Delta C})$.

And finally, Fig. \ref{figure:Editorcites}(C) combines three editor service measures,    editorial power ($N_{E}$), longitudinal trends of editorial quality judgement ($\beta_{\tau}$), and editorial citation renumeration ($C_{E}$) in a single scatter-plot visualization. First,  each point, corresponding to an editor, is colored according to the trend in $z_{A}$ over time. To be specific, we used ordinary least squares regression to estimate the temporal trend in the linear model $z_{A,\tau} = \beta_{0}+\beta_{\tau} \tau$,  captured by the coefficient $\beta_{\tau}$,  using all the articles with $t\leq 2014$ for a single editor. For the editors with significant positive (negative) $\beta_{\tau}$ (using the significance level $p<0.1$) we colored their datapoint cyan (orange); we observed 194 editors with a significant positive trend and 100 with a significant negative trend. Remarkably, two top-10 editors  -- Vladimir Uversky (rank $r=1$) and Matjaz Perc ($r=7$) -- are distinguished not only by their power and total citation renumeration, but also by the fact that the trends in the citation impact of articles they have accepted has been negative over time ($\beta_{\tau}<0$). However, if we eliminate these top two editors (according to $N_{E}$) from the orange and cyan subsets, and calculate the power-law relationship between renumeration and editorial service ($C \propto N^{\gamma}$), we obtain nearly identical fits (shown as solid colored lines) across the entire range of  $N_{E}$,  with approximately equivalent scaling exponent: $\gamma = 0.41$ (orange) and 0.43 (cyan).

\section{Summary and discussion}

We analyzed the largest journal in the world, focusing on quantifiable measures of  editorial power and decision processes. 
Given its size, the impact that PLOS ONE has on the production of scientific knowledge is high, and given the implicit constraints in monitoring and managing  such a large complex system, several of our findings are worrisome, with plausible explanations that range from editorial apathy to misconduct in their service to science.  Thus, our study  demonstrates why large megajournals should record, monitor, and embrace transparency, and to make sure the incentives for editor service do not introduce unintended conflicts-of-interest. As science continues to grow, these conflicts-of-interest may become more difficult to avoid, e.g. in large teams or a large journal, because the difficulty in monitor individual activities  may foster the conditions for misconduct \citep{petersen_quantitative_2014}. 
 
By analyzing the entire editorial board comprised of nearly 7,000 editors, we revealed the  remarkable levels of inequality in the activity levels among  PLOS ONE editors.  While the Gini-index equal to 0.58 we calculated for editorial power is remarkably high, it is nevertheless smaller than the Gini index calculated for university funding  (Gini index exceeding 0.7 )\citep{ScienceInequality}  and career achievement (Gini-index for total researcher citations ranging from 0.6 to 0.8)  \citep{CumAdvEPJDS}. A corollary to such power disparity is the additional external power, prestige, and familiarity it may lead to, providing an additional mechanism contributing to cumulative advantage in science \citep{DeSollaPrice_bibliometric_1976,BB2,CumAdvEPJDS}.

To get a better idea of the power levels among the most prolific editors, we inspected the top-100 editors, ranked according to the total number of articles $N_{E}$, and observed a wide range of mean acceptance times among them, ranging from $\Delta_{E}=175$ to as short as $\Delta_{E}=56$ days, on average. Not surprisingly, and not upstanding, several of the editors with the shortest $\Delta_{E}$ were among the top-10 editors. For example, on average, articles edited by the most active editor, Vladimir Uversky, appear every 3.2 days, a feat which is partly attributable to the relatively short mean article acceptance time $\Delta_{E}=77$ days (the editor average is 130 days). The variability in acceptance time within each editor's  profile was also high (see Fig. \ref{figure:Figure-S1}A). One potential explanation for the prevalence of such short review times is the portability of external reviews, which can be used in the PLOS ONE editorial decision process \citep{Bjork_Megajournalgrowth_2015}.

Assuming that editors have fixed time resources, this  power disparity directly translates into a disparity in the time editors (appears to) spend in evaluating each article. The results of our citation model  in Section \ref{subectionmodel1} provide supporting evidence that the quality of his/her service diminishes significantly with higher work-load, as editors' ability to evaluate article quality diminishes significantly over time.  
While it is clear from this statistical analysis that some editors are overactive, it is important to note that this  is not by itself a certain indicator of  misconduct. 
Indeed, short of a case-by-case analysis requiring additional review data,  distinguishing good intentions from bad inevitably leads to an in-between gray area  where it is difficult to distinguish. Nevertheless,  we leveraged the longitudinal aspect of the data using two fixed-effect models  to identify trends within editorial profiles.

In the first model, we used the wide range of variation in editor activity as well as longitudinal time to identify 
 factors that can explain the variation article quality -- proxied by its citation impact. We used a normalized citation measure to overcome systematic measurement biases that are prevalent in citation analysis.
Our results show  that editors  tend to be biased in favor of authors with whom they have had previous editorial experience. This feature alone is not necessarily unexpected, given the strength of social ties underlying other aspects of the scientific endeavor \citep{petersen_quantifying_2015}.
However, we found that this effect was pronounced among the top-10 editors. In all, these results of this citation model are particularly worrisome, showing  the ability  of editors to evaluate research quality decreases over their editorial career and that editors exhibit bias in favor of  repeat authors -- a signature of ``backscratching'' behavioral bias across social ties -- which may also be related to poorer quality of the articles  accepted. 

In the second model, we looked for social factors that might explain the wide variation in acceptance times.
We found that articles that have a larger fraction of citations citing the editor's research were likely to be accepted faster. 
The most plausible reason that an author might cite the editor's work is because PLOS ONE editors are acting scientists, and thus central figures within their community. Moreover, as the editor continues to publish him/herself, then clearly he/she has more publications to cite over time.
Nevertheless, we found that this new type of ``self-citation''  \citep{Fowler_Self-citation_2007,Costas_Self-citations_2010} had a more prominent effect  
when considering the presence of repeat authors, as  captured by the indicator variable $R_{A,E}=1$ which identifies the articles containing at least one author that published two or more times with that given editor. Figure \ref{figure:ModelMarginalEffects}(B) shows a stronger marginal effect of reduced acceptance time when the article does not contain any repeat authors.
This effect is related to the literature on self-citations, which offers various plausible explanations in addition to  citation rigging, e.g. signaling prestige in cross-disciplinary mobility \citep{Hellsten2007} as well as bias towards citing one's past collaborators, to explain the  self-citation frequency which ranges between 20 to 40\% \cite{Costas_Self-citations_2010}. 
Here, however, since the editor has decision power, the incentives to sway that decision are more clear, providing further evidence that directed citations are indeed an effective form of renumeration \citep{Fowler_Self-citation_2007}.

If editor's are truly privy -- and responsive -- to such subtle payments,  the follow-up question is how much could one really gain by playing this game?
Because it is extremely difficult to measure and interpret the context behind individual citations, we appealed to the large data size to look for evidence of variation in editor citation rates ($f_{E}$) depending on whether the citations arise from repeat authors or not. Figure \ref{figure:Editorcites} provides substantial evidence that a small but significant number of editors may be aware of this game, and our estimates of  the total citations attributable to $\Delta R$ place a lower bound on this citation renumeration in the hundreds of citations. While editors certainly deserve credit for their service to science, it is important to address the possibility that some editors may have  more covert intentions underlying their excess editorial activity.  

 \section{Conclusions}
 
 Despite the plethora of researcher metrics  being developed and refined in order to evaluate individual scientific achievement \citep{TheMetricTide}, much less attention has been paid to the scientific actors at the other end of the negotiation table -- the editors who serve as scientific gatekeepers. Holding the power to accept or reject a scientific article has obvious ramifications for the authors of the manuscript. Even more, this decision  has the profoundly determines whether or not  their conclusions enter in to the corpora of scientific literature, and thus, the cannon of scientific knowledge. 

As such, it goes without saying that the power that editors hold over the body of scientific knowledge comes also with a lot of  responsibility. Maintaining certain levels of quality is crucial for the advancement of knowledge, which is often incremental, but nevertheless  cumulative, building and relying on what has been previously published, and to some extent,  what has been legitimized by the scientific review process. By allowing publication standards to diminish, and possibly even disappear, science may become susceptible to incorrect, if not fraudulent, mis-knowledge. With the democratization of online publication in the general sense, it is not implausible that the same prevalence and ease of  misinformation spreading in the digital publication world, where the role of editors has been largely diminished or eliminated, could also develop within science. Thus, it is important to highlight the value and responsibility that editors have in maintaining the integrity of science.

We conclude with some editorial policy recommendations. First, especially in the case of mega-journals featuring a distributed editor system comprised of acting researchers, we encourage the journals to go beyond the example of PLOS ONE in linking editors to published articles, and to actually record and evaluate  editors' activity levels. This is partially a transparency issue, but also a responsibility and sanctioning issue that is necessary for the management of science journals. As gatekeepers to our knowledge base, science editors have a distinct responsibility in remaining unbiased. Our results, however, indicate that social biases are nevertheless predominant features despite best efforts. Second, electronic-only journals which do not have volume restrictions should nevertheless consider placing restrictions on the number of articles an editor can oversee at a time and per year. In addition to discouraging editors from taking advantage of their power, it  would also encourage higher quality standards for accepting an article for publication. By implementing such  editorial policy changes at PLOS ONE, it  would certainly make for an interesting policy experiment, providing an additional opportunity to observe shifts in editorial behavior, and  possibly strengthening the case for tying the observed behavioral trends to outright misconduct. 

\section*{References}

\bibliography{Policy}
\bibliographystyle{model2-names}

\clearpage
\newpage

\begin{table}[b!]  
\caption{Results of a fixed-effects model for which the dependent variable is the citation impact of an individual article ($z_{A}$).  Model parameters estimated using editor, year, and subject-area fixed-effects. Only editors  with $N_{A}\geq 10$  articles for the  years $t\leq2014$ are analyzed; see Eq.~\ref{Eqn1} for the full model specification. 
Estimates calculated using robust standard errors. The third and fourth columns show estimates including interaction effects.} 
\resizebox{0.99\columnwidth}{!}{ 
\def\sym#1{\ifmmode^{#1}\else\(^{#1}\)\fi}
\begin{tabular}{l*{4}{cc}}
\hline\hline
            &\multicolumn{8}{c}{Detrended citation impact, $z_{A}$}         \\
            &\multicolumn{2}{c}{Model coeffecient}           &\multicolumn{2}{c}{Standardized coeff.($\hat\beta$)}           &\multicolumn{2}{c}{ $R \times \ln \tau$ }           &\multicolumn{2}{c}{$T_{10} \times R \times \ln \tau$}           \\
\hline
$\ln k_{A}$  &       0.285\sym{***}&     (0.000)&           0.161\sym{***}           &     (0.000)       &       0.286\sym{***}&     (0.000)&       0.286\sym{***}&     (0.000)\\
$\ln \Delta_{A}$ &     -0.127\sym{***}&     (0.000)&     -0.0755\sym{***}                &     (0.000)       &      -0.127\sym{***}&     (0.000)&      -0.127\sym{***}&     (0.000)\\
$\ln \tau_{E}$ &     -0.178\sym{***}&     (0.000)&        -0.143\sym{***}             &    (0.000)        &      -0.174\sym{***}&     (0.000)&      -0.175\sym{***}&     (0.000)\\
$R_{A,E}$ &       0.0895\sym{***} &     (0.000)&      0.0895\sym{***}&     (0.000)   &       0.107\sym{***}&     (0.000)&       0.106\sym{***}&     (0.000)\\
 $R_{A,E} \times \ln \tau_{E}$ &                     &            &                     &            &     -0.0320\sym{**} &     (0.005)&     -0.0252\sym{*}  &     (0.028)\\
$T_{10,E} \times \ln \tau_{E}$ &                     &            &                     &            &                     &            &     0.0445         &     (0.433)\\
$T_{10,E} \times R_{A,E} \times \ln \tau_{E}$ &                     &            &                     &            &                     &            &   -0.103\sym{*}  &     (0.017)\\
Constant      &      -0.813\sym{***}&     (0.000)&      -0.970\sym{***}&     (0.000)&      -0.823\sym{***}&     (0.000)&      -0.818\sym{***}&     (0.000)\\
\hline
Dummy for $year$ & y & & y &  & y & & y &\\
Dummy for $SA$ & y & & y & & y & & y & \\
\hline
\(N\)       &      102741         &            &      102741         &            &      102741         &            &      102741         &            \\
adj. \(R^{2}\)&       0.035         &            &       0.035         &            &       0.035         &            &       0.035         &            \\
F           &       182.7         &   (0.000)          &       182.7         &       (0.000)       &       173.4         &    (0.000)          &       150.0         &     (0.000)         \\
$df_{\tt model}$         &          16         &            &          16         &            &          17         &            &          20         &            \\
$df_{\tt clusters (E)}$    &        3084         &            &        3084         &            &        3084         &            &        3084         &            \\
\hline\hline
\multicolumn{9}{l}{\footnotesize \textit{p}-values in parentheses}\\
\multicolumn{9}{l}{\footnotesize \sym{*} \(p<0.05\), \sym{**} \(p<0.01\), \sym{***} \(p<0.001\)}\\
\end{tabular}
}
\label{table:reg1} 
\end{table} 

\begin{table}[b!]  
\caption{ Results of a fixed-effects model for which the dependent variable is the logarithm of the acceptance time for an individual article ($\ln \Delta_{A}$).  Model parameters estimated using editor, year, and subject-area fixed-effects. Only editors  with $N_{A}\geq 10$ are analyzed; see Eq.~\ref{Eqn2} for the full model specification.  
 Estimates calculated using robust standard errors. } 
\resizebox{0.99\columnwidth}{!}{ 
\def\sym#1{\ifmmode^{#1}\else\(^{#1}\)\fi}
\begin{tabular}{l*{2}{cc}}
\hline\hline
            &\multicolumn{4}{c}{Acceptance time model, $\ln \Delta_{A}$}         \\
            &\multicolumn{2}{c}{Model coeffecient}           &\multicolumn{2}{c}{Standardized coeff.($\hat\beta$)}           \\
\hline
$z_{A}$     &     -0.0343\sym{***}&     (0.000) &     -0.0343\sym{***}&     (0.000)\\
$\ln k_{A}$  &      0.0529\sym{***}&     (0.000) &      0.0297\sym{***}&     (0.000)\\
$f_{A}$ &      -0.674\sym{***}&     (0.000) &    -0.00895\sym{***}&     (0.000)\\
$\ln \tau_{E}$ &       0.170\sym{***}&     (0.000) &       0.137\sym{***}&     (0.000)\\
$R_{A,E}$ &     -0.0878\sym{***}&     (0.000)&     -0.0878\sym{***}&     (0.000)\\
constant     &       3.976\sym{***}&     (0.000)&       4.152\sym{***}&     (0.000)\\
\hline
Dummy for $year$ & y & & y &  \\
Dummy for $SA$ & y & & y &  \\
\hline
\(N\)        &      128,734         &            &              &            \\
adj. \(R^{2}\)&           0.064         &            &             &            \\
F          &      255.7         &  (0.000)          &         &          \\
$df_{\tt model}$        &          18         &            &                  &            \\
$df_{\tt clusters (E)}$      &        3,748         &            &                &            \\
\hline\hline
\multicolumn{5}{l}{\footnotesize \textit{p}-values in parentheses}\\
\multicolumn{5}{l}{\footnotesize \sym{*} \(p<0.05\), \sym{**} \(p<0.01\), \sym{***} \(p<0.001\)}\\
\end{tabular}
}
\label{table:reg2} 
\end{table} 

\clearpage
\newpage


\begin{center}
{\large \bf Supplementary Information}
\end{center}

\bigskip
\renewcommand{\theequation}{S\arabic{equation}}
\renewcommand{\thefigure}{S\arabic{figure}}
\renewcommand{\thetable}{S\arabic{table}}
\renewcommand{\thesection}{S\arabic{section}}
 
\setcounter{equation}{0}  
\setcounter{figure}{0}
\setcounter{table}{0}
\setcounter{section}{0}
\setcounter{page}{1}

\bigskip
\begin{center}
{\large \bf Quantifying the distribution of editorial power and manuscript decision bias at the mega-journal PLOS ONE} \\
\bigskip
Alexander M. Petersen$^{1}$ \\
\bigskip
$^{1}$Management Program, School of Engineering, University of California, Merced, California 95343 \\
\bigskip
\end{center}
\bigskip

\section{Supplementary Methods}

\subsection{Name disambiguation problem among editors and authors}
\label{NameDisamb}
Due to the name disambiguation problem -- i.e. it is difficult to distinguish common last name and first name initial combinations in TRWOS data -- there are certain abbreviated name combinations that we ignored in aspects of our analysis. First, in order to 
determine if an article was coauthored by a PLOS ONE editor, there were certain editor names which were too similar in their abbreviated forms,  e.g. Shree Singh and and Seema Singh, who both occur in TRWOS records as ``Singh S''. Thus, for those editor name abbreviations which have a degeneracy of 2 or greater, we do not count articles with these abbreviated names as being coauthored by an editor.

Second, this name disambiguation problem occurs in the identification of the top authors within the article set of each editor. Thus, using the editor name degeneracy set as our baseline, we also ignored all surnames -- independent of first name initial -- for the common PLOS ONE editor list. As such, the list of common surnames ignored in the coauthor analysis are :  Singh, Isalan, Hoheisel, Lo, Castresana, Liu, Zheng, 
Yang, Deb, Qiu, Chang, Zhou, Bhattacharya, Tang, Lee, 
Xu, Li, Cheng, Wang, Scott, Yu, Tan, Miao, 
Williams, Klymkowsky, Kaltenboeck, Zhang, Chen, He, 
Song, Brown, Lin, Brody, Wei, Kumar, Yan, Shi, 
Carvalho, Rogers, Ng, Ray, Phillips, Soriano-Mas, Paul, 
Fox, Butler, Ma, Wu, Carter, Xie, Hector, Wright, 
Caldwell, Fang, Sorensen, Lam, Chan, Stewart, Huang, 
Gravenor, Pan, Gupta, Smith, Lu, Cao, Xia, Ho, 
Moore, Liang, Franco, Parida, Zhao, Wilson, Gilbert, 
Nigou, Redfield, Paci, Park, Sun, Zhu, Chalmers, 
Clark, Colombo, Zuo, Das, Tian, Moreno, Meng, Gray, 
Schweisguth, Lopez-Garcia, Yue, Johnson, Wong, Medina, 
Fung, Kato, Roberts, Hwang, Hsieh, Wen, Knight, 
Csernoch, Anderson, Grant, Clarke, Jiang, Jones, Rao, 
Feng, Nguyen, Choi, Thomas, Chiu, Samuel, Gordon, 
Heutink, Evans, Martin, Ren, Berger, Kim, Han, Mao, 
White, McCutcheon, Temussi, Taylor, Schmitt, Kerby, 
Miller, Roy, Pereira, Shankar, Aoki, Jackson, Adams, 
Russell, Thompson, Abe, Duan, Hong, Borras, Costa, 
Yam, Porollo, Stumbles, Agarwal, Beier, Xiao, Beaudoin, 
Nosten, Shen, Feldman, Hall, Raible, Yin, Kelly, 
Simos, Knudsen.

\begin{figure*}
\centering{\includegraphics[width=0.99\textwidth]{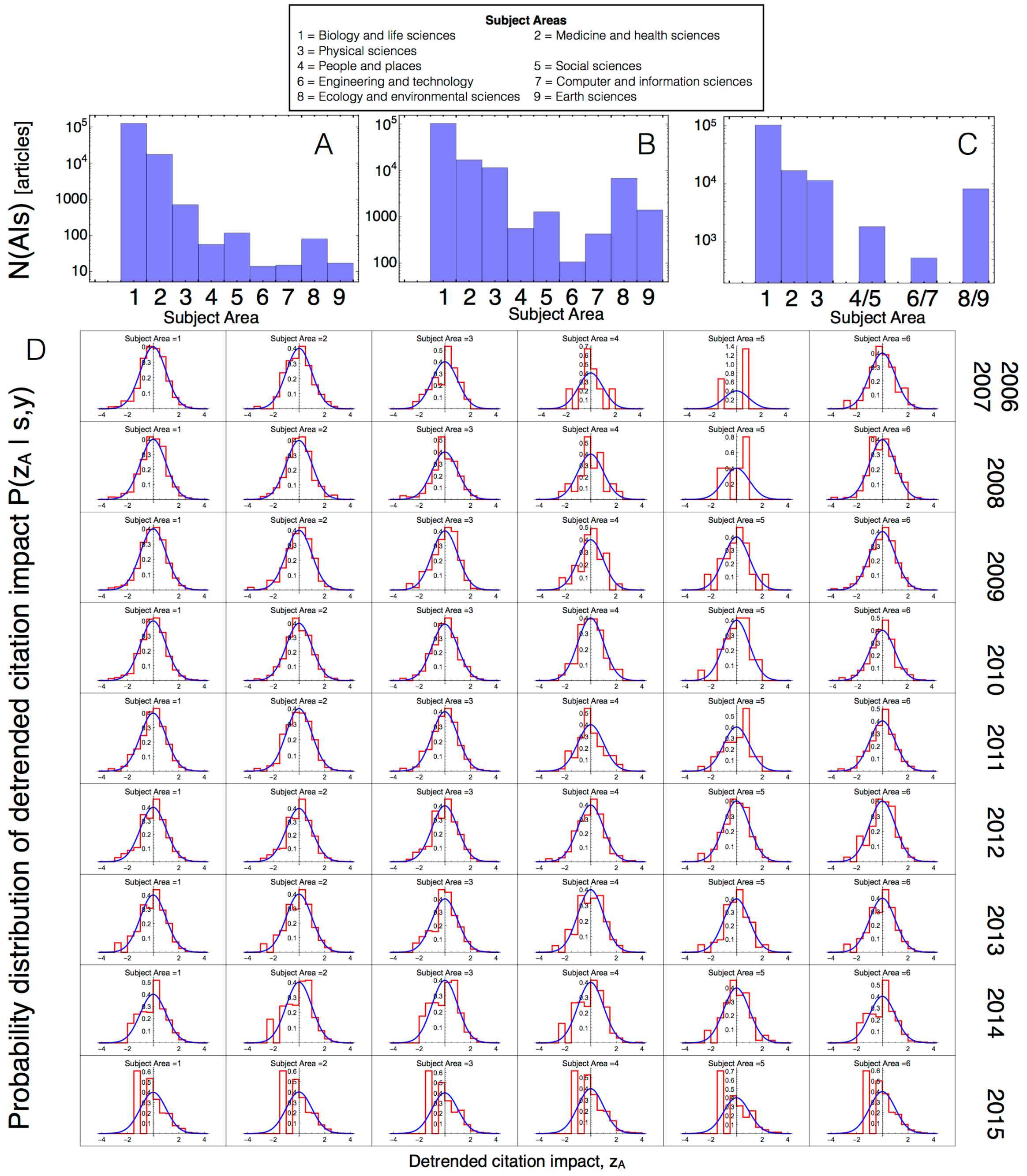}}
\caption{\label{figure:Figure-S1}   {\bf Distribution of article characteristics: subject area and detrended citation impact.} (A) Count distribution of the number of articles by principal subject area (no articles were observed  with ``Science policy''  as the principle SA). (B) Count distribution by SA after applying redistribution rule that if the principal SA=1, then use the SA with the second-highest weight. (C) Count distribution by SA after merging into 6 refined subject areas, which are used throughout the analysis. (D) Empirical probability distribution $P(z|s,t)$ for each SA and year combination (red bins) and baseline normal distribution $N(0,1)$ (blue curve) shown to demonstrate the time-independence of the normalized citation impact variable.  Since all 2006 articles were published in December, we merged these publications with 2007. The only articles with poor convergence to the $N(0,1)$ distribution are the relatively recent 2015 articles;  these publications are not included in the regression analysis.}
\end{figure*}

\begin{figure*}
\centering{\includegraphics[width=0.59\textwidth]{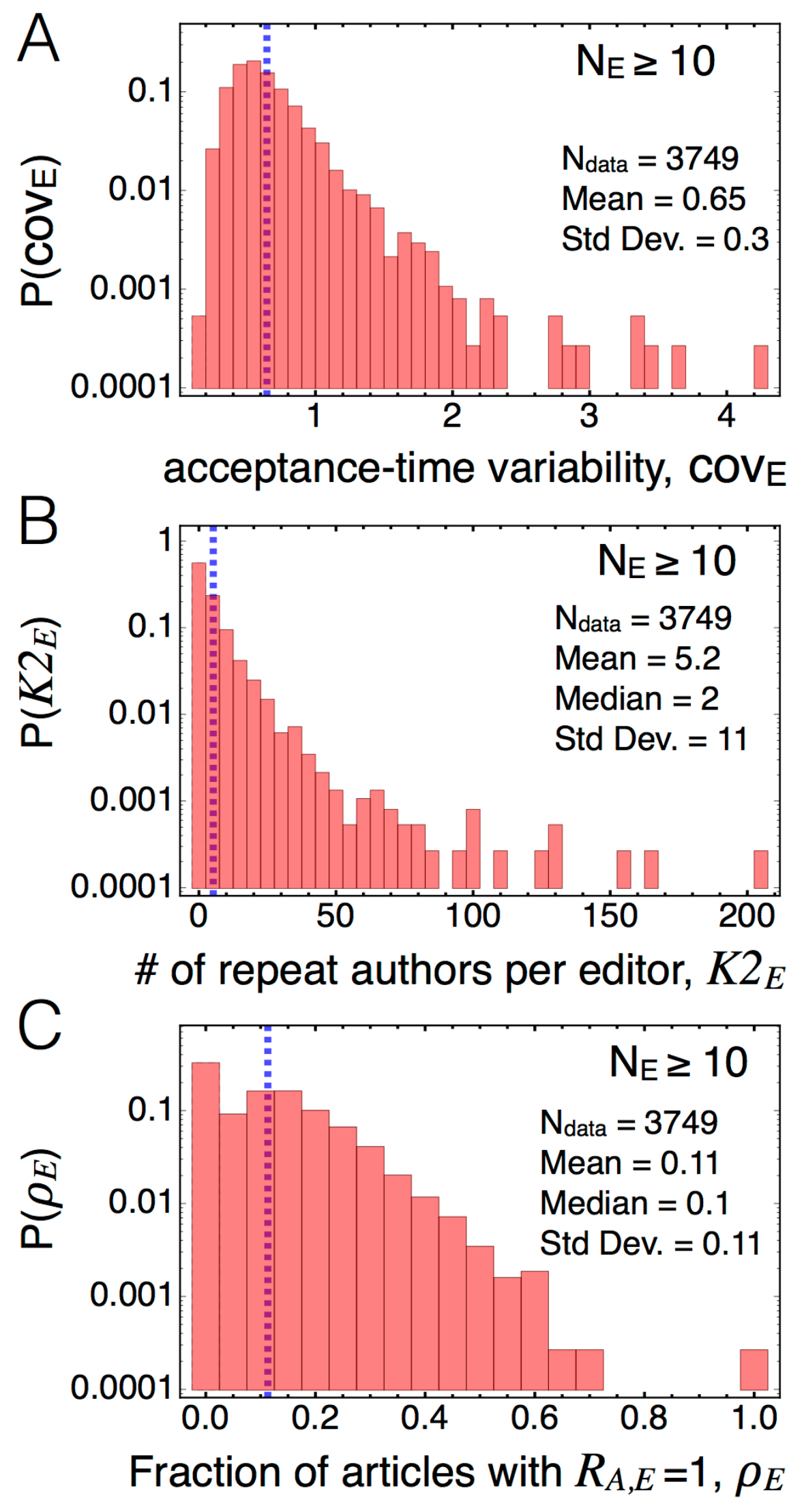}}
\caption{\label{figure:Figure-S2}   {\bf Editor-level article characteristic distributions.} (A) Probability distribution $P(cov_{E})$ of the variability in $\Delta_{E}$ expressed as the coefficient of variation (the ratio of the standard deviation of $\Delta_{E}$ normalized by the mean $\Delta_{E}$ for a given editor). For most editors the mean $\Delta_{E}$ is rather characteristic, however some editors show a wide range of variability. (B) Probability distribution   $P(K2_{E})$ of the number $K2_{E}$ of repeat authors, i.e. the authors that have appeared on 2 or more of the $N_{E}$ articles within a given editor's article set. (C) Probability distribution   $P(\rho_{E})$ of the fraction $\rho$ of the total articles of a given editor featuring a repeat author (i.e. fraction of articles with indicator value $R_{A,E}=1$). In each panel we only analyze editors with $N_{E}\geq 10$ articles to avoid small sample size fluctuations; vertical dashed lines indicate distribution mean.}
\end{figure*}


\end{document}